\def\be{\begin{equation}}
\def\ee{\end{equation}}
\def\bea{\begin{eqnarray}}
\def\eea{\end{eqnarray}}
\def\bse{\begin{subequations}}
\def\ese{\end{subequations}}

\documentclass[prl,twocolumn,showpacs,preprintnumbers,amsmath,amssymb]{revtex4}
\usepackage{graphicx}
\usepackage{dcolumn}
\usepackage{bm}

\begin{document}

\title{
Edge Transport in 2D Cold Atom Optical Lattices
}
\author{V. W. Scarola and S. Das Sarma }
\affiliation{Condensed Matter Theory Center, Department of
Physics, University of Maryland, College Park, MD 20742}

\date{\today
}
\begin{abstract}
We theoretically study the observable response of edge currents in two dimensional 
cold atom optical lattices.  As an example we use Gutzwiller mean-field theory to relate persistent 
edge currents surrounding a Mott insulator in a slowly rotating trapped Bose-Hubbard system 
to time of flight measurements.  We briefly discuss an application, the detection 
of Chern number using edge currents of a topologically ordered optical lattice insulator.
\end{abstract}


\maketitle

The response theory of transport is a remarkably precise 
framework used to analyze the observable effects 
of applied potentials in a broad class of solid state systems.  It is natural to ask how 
experiments on neutral cold atoms confined to optical lattices \cite{Oplat}, 
predicted to hold a variety of novel phases of matter \cite{Lewenstein} 
similar to those found in the solid state, can 
make contact with an equivalent quantitative framework.  Recent experimental work on cold atom optical lattices 
demonstrates essential ingredients in establishing quantitative response: 
applied potentials and detection of conserved quantities.  A primary 
tool for detection relies on time of flight (TOF) imaging which provided the first evidence 
for Bose-Einstein condensation \cite{Ketterle} and revealed phases of  
optical lattice realizations of Bose-Hubbard (BH) models \cite{Jaksch,Greiner,Altman,Bloch,Spielman}.  

By combining TOF with externally applied potentials recent work 
has demonstrated transport in one dimensional optical lattices 
\cite{Pezze,Porto}.  In a closed two dimensional system the notion 
of ``transport'' is less direct.  A recent 
experiment has applied rotation to weak lattices \cite{Cornell} confining 
bosons.  While 
far from the single band BH limit, this experiment reveals vortex pinning 
arising from the weak lattice.  Recent work \cite{magnetic,Demler_FQHE} also 
suggests that uniform effective 
magnetic fields (equivalent to rotation) may be applied to optical lattices 
already in the BH limit.  Either implementation, rotation or an effective magnetic 
field, can be used as an applied potential valuable in establishing persistent 
currents, and therefore transport, in two dimensional lattices.     

Concurrent with experimental progress, a variety of cold atom phases have been 
proposed in two dimensional optical lattices \cite{Lewenstein}.  Some of 
the proposed lattice models have rich phase diagrams with particularly intriguing 
or even unknown ground states, including:  extended BH  
models \cite{Goral,Lewenstein,Scarola}, higher band spin models \cite{Girvin}, fractional quantum 
Hall models \cite{Demler_FQHE}, and the 
Kitaev spin model \cite{Kitaev,Demler_Spin,Zoller}.  We ask how 
insulating phases arising in two dimensional lattice models  
can be studied using a combination of externally applied potentials and TOF.    
 
Below we argue that trapping leads to edge states which serve as a probe of bulk 
insulating states.  As a concrete and relevant  
example we study the slowly rotating BH model in detail.  Other 
studies have considered vortex configurations in the superfluid phase of the rotating uniform      
BH model \cite{Wu,Carr}.  Here we study edge effects in the Mott insulating 
phase of the slowly rotating trapped BH model.  We propose that diamagnetic response of  
edge states can indeed be observed thereby offering a quantitative response probe of a variety of bulk 
two dimensional insulators.  We briefly discuss implications for another insulator 
where edge states can be used to detect the 
Chern number \cite{Hatsugai,Kitaev,Qi} of a topologically ordered insulator, the non-Abelian ground state of the Kitaev model.

We first note that response to externally applied fields can be 
obtained at a quantitative level by analyzing TOF measurements.     
TOF can be related to the momentum density, $\rho_{\bm k}$, of 
particles with lattice momentum ${\bm k}$ originally trapped in an optical lattice.  Observation 
of $\rho_{\bm k}$ (with sufficient accuracy) can be combined with input parameters to restore 
quantities of the form:
$
\mathcal{J}\equiv\sum_{{\bm k}} \mathcal{W}_{\bm k} \rho_{\bm k},
$
where $\mathcal{W}_{\bm k}$ is any function of ${\bm k}$ which can be accurately 
determined from input experimental parameters.  
By defining $\mathcal{W}_{\bm k}= \mathcal{M}_{\bm k}(\partial E_{\bm k} /\partial k_{\alpha})$ 
we obtain two examples: the free-particle number and energy currents 
in the direction $\alpha$ with the choices 
$\mathcal{M}_{\bm k}=1$ and $\mathcal{M}_{\bm k}=E_{\bm k}$, respectively.  
Here, $E_{\bm k}$ is the single particle energy determined by 
optical lattice parameters.  As we will show, diamagnetic current flowing 
along sufficiently narrow edges of optical lattice insulators can be written in the form 
$\mathcal{J}$ allowing restoration of the edge current.

To study the observable response of insulating states in trapped optical lattices 
we consider the two-dimensional BH model on a square lattice in the presence 
of rotation (or, equivalently, an effective magnetic field \cite{magnetic,Demler_FQHE})
as a first step in establishing quantitative response in systems  
nearest ongoing experiments.  Using the Peierls 
substitution \cite{Peierls} the BH model in the rotating frame is:
\begin{eqnarray}
H&=&-t\sum_{\langle i, j\rangle}
\left[\exp{(i\mathcal{A}_{i,j})}a_{i}^{\dagger}a_{j}^{\vphantom{\dagger}}
+ {\rm{h.c.}}\right]
\nonumber \\
&+& \frac{U}{2}\sum_{i}n_{i}(n_i-1) 
- \sum_{i}\left(\mu - \tilde{\kappa}\vert {\bm r}_i \vert^2 \right)n_{i},  
\label{Hamiltonian}
\end{eqnarray}
where $a_{i}^{\dagger}$ and  $n_i$ are the boson creation and number 
operators at the site $i$, respectively.  The 
parameters include the hopping, $t$, the onsite interaction energy, $U$, and the chemical potential, $\mu$.  The last term is due to the trapping potential which adds a site dependent chemical 
potential at the square lattice coordinate in the $xy$ plane, $\bm{r}_i=(i_x,i_y)$,  
parameterized by a modified trapping parameter $\tilde{\kappa}=\kappa-m(\Omega a)^2/2$, in units of the 
lattice spacing, $a$ (half the wavelength of the lasers defining the optical lattice).  The trapping parameter 
is modified by a term due to rotation, with angular frequency, $\Omega$, of particles of mass $m$.  In what 
follows we find that, for $\text{Rb}^{87}$ atoms with $\mu/U=0.4$, $U/E_R\sim 0.1$, and 
$\kappa/U=1.2\times10^{-3}$ the modification is  
negligible for the rotation frequencies studied here 
giving $\tilde{\kappa}\approx\kappa$.  $E_R=h^2/(8ma^2)$ is the recoil energy.    
The rotation also modifies the hopping term to give a phase:
$
\mathcal{A}_{i,j}=(\pi^2\hbar\Omega/2E_R)\int_i^{j} (\hat{z}\times{\bm r})\cdot d{\bm s}, 
$
which can be thought of as an integral over a vector potential due to an effective 
magnetic field, $B^*\hat{z}$, acting on an effective charge $q^*$ such that $q^* B^* =2m\Omega $.  In the 
rotating frame the neutral bosons experience an effective magnetic field which induces a  
persistent current opposing the applied effective field.


In the linear response regime,     
$
\hbar \Omega/E_R \ll 1,
$
we can define \cite{Scalapino} 
a number current in the direction $\alpha$ in response to the static applied potential, $\mathcal{A}$: 
$
J_{i,\alpha}= it\hbar^{-1}\left(a_{i}^{\dagger}a_{i_{\alpha}}^{\vphantom{\dagger}}- {\rm{h.c.}}\right)
-\hbar^{-1} \mathcal{A}_{i,i_{\alpha}}K_{i,i_{\alpha}}
$
where the first term is the paramagnetic number current and 
the second term is the diamagnetic number current which contains the kinetic term:
$
K_{i,i_{\alpha}}\equiv t\left(a_{i}^{\dagger}a_{i_{\alpha}}^{\vphantom{\dagger}}+ {\rm{h.c.}}\right).
$
The total diamagnetic number current is an observable response to 
our externally applied field giving:
$
J^{\rm D}_{\alpha}=-\hbar^{-1}\sum_{l}K_{i,i_{\alpha}}\mathcal{A}_{i,i_{\alpha}}.
$
In general $J^{\rm D}$ cannot be written in the form $\mathcal{J}$ and it is therefore not 
clear how we can relate such a quantity to TOF measurements.  In what follows we will 
show that when the diamagnetic current is confined to the edge of the system (and therefore $\mathcal{A}$ 
varies sufficiently slowly) we {\em can} write an approximation to $J^{\rm D}$ which in turn can 
be related to TOF.  Consider the following approximation:
$
\bar{J}^{\rm D}_{\alpha} \equiv -\hbar^{-1}\mathcal{A}_{i^e,i^e_{\alpha}}\sum_{i}K_{i,i_{\alpha}}, 
$
where:
$
\mathcal{A}_{i^e,i^e_{\theta}}=(\pi^2\hbar\Omega/2E_R) (\hat{z}\times {\bm r}_{i^e})\cdot \hat{\theta}.
$
Here 
$
{\bm r}_{i^e}
$
indicates the average position of the edge superfluid order parameter,  
$
\psi_i=\langle a^{\dagger}_{i} \rangle,
$
 giving 
$
\hat{r}\cdot{\bm r}_{i^e}= 
(\sum_{i}\vert \psi_i \vert ^2  \hat{r}\cdot{\bm r}_i )
/(\sum_{i}\vert \psi_i \vert^2).
$ 
By Fourier transforming and taking the expectation value with respect to the ground state we find:
$
\langle \bar{J}^{\rm D}_{\theta} \rangle 
= -2t\mathcal{A}_{i^e,i^e_{\theta}}(\hbar Z)^{-1} 
\sum_{{\bm k},{\bm \delta}} \rho_{{\bm k}} \cos \left({\bm k}\cdot{\bm \delta} \right), 
$
where $Z$ is the number of nearest neighbors 
with lattice vectors ${\bm \delta}$.  We now have a quantity 
written in terms of the lattice momentum distribution:  
$
\rho_{\bm k}= \sum_{i,j} \exp(-i {\bm k}\cdot({\bm r}_i-{\bm r}_{j})) 
\langle  a^{\dagger}_i a^{\vphantom{\dagger}}_{j}\rangle,
$
which, we assert, yields an accurate measure of the diamagnetic current provided  
the current flows along the edge.  Our assertion can be written:
$
\langle J_{\theta}^{\rm D} \rangle_{e} \approx \langle \bar{J}_{\theta}^{\rm D} \rangle_{e}, 
$
where $\langle \rangle_e$ indicates averaging in a ground state with only edge current.  As we will see 
this relation allows us to probe the edge flow around bulk insulators in optical lattices 
but does not necessarily hold for the rotation of a bulk superfluid in a trap.  To 
continue with our example of the rotating BH model we relate 
$\langle \bar{J}^{\rm D}_{\theta} \rangle$ to an observable TOF signal.   

TOF signal can be 
directly related to the momentum distribution even in a slowly rotating optical lattice.  In the 
following we assume that the particles do not interact after release from the trap.  
We may then apply the free particle propagator $K_{\rm p}$ to 
a single particle Bloch state in the rotating frame, $\phi_{{\bm k}}({\bm r})$,  initially 
confined to the lattice.   We project it onto a imaged 
state $\Phi^s$ with imaged coordinates ${\bm r}_s$ in the laboratory frame. 
For slow rotation we find:
$
\Phi^s({\bm r}_s)=\int K^{\rm p}({\bm r};{\bm r}_s) 
\phi_{{\bm k}}({\bm r})  d{\bm r}
\propto \delta^{\prime}_{{\bm k},{\bm Q}}
\vert \tilde{w}\left( {\bm k} -\Delta t \Omega \hat{z}\times{\bm k} \right) \vert^2,
$
where $\Delta t$ is the time taken to propagate from the lattice to the imaging screen, 
$\delta^{\prime}$ indicates equivalence up to a reciprocal lattice vector, and $\tilde{w}$ is the Fourier 
transform of the non-rotating Wannier function.  
Here the lattice wavevector gets mapped to position on the screen in free particle propagation:
$
{\bm Q}({\bm r}_s)=(m {\bm r}_s)/(\hbar \Delta t).
$
We have derived the above expression to lowest order in $
(\hbar \Omega/E_R)^2,
$
consistent with our linear response approximation.  
The imaged total density is then: 
$
n^s({\bm r}_s)\approx \left ( m/(\hbar \Delta t) \right )^2 
\rho_{{\bm Q({\bm r}_s)}} \vert \tilde{w}\left( {\bm Q({\bm r}_s)} 
-\Delta t \Omega\hat{z}\times{\bm Q({\bm r}_s)} \right) \vert^2. 
$
We have found, as in the non-rotating case \cite{Grondalski,Svistunov}, that up to an overall 
Gaussian-like function, $\tilde{w}$, the imaged density on the screen gives $\rho_{\bm k}$.  
We now study the slowly rotating BH model 
under the assumption that $\rho_{\bm k}$ can be accurately extracted from measurements.

We calculate the ground state of the rotating BH model using a modification of the Gutzwiller 
mean-field ansatz \cite{Rokhsar,Jaksch}.  We assume a product state in the 
Fock number basis $\vert N_i \rangle$, of the form:
$
\Psi=\prod_{i}\sum_{N_i=0}^{N_c} f^{i}_{N_i} \vert N_i \rangle,
$
where the $(N_c+1)N$ complex variational parameters $f^{i}_{N_i}$ are chosen to minimize the 
ground state energy of $H$ on $N$ lattice sites.  In what follows we choose $N=50\times 50$ where 
the confinement forces the atoms to occupy no more than $\approx 45\times 45$ sites.  We also find that 
$N_c=5$ gives suitable convergence for the low chemical potentials 
studied here.  We minimize $\langle H \rangle$ using 
the conjugate gradient method.  To treat large systems we have developed a three step 
minimization procedure with a computational cost that scales linearly with $N$.  Using our product ansatz we first 
find the ground state assuming that each site 
is an independent system with $\Omega=0$.  In our second step we minimize the energy of 
the whole system using step one as an initial guess, while keeping $\Omega=0$.  
This step shows \cite{Zakrzewski} excellent agreement with Monte Carlo simulations \cite{Svistunov}.  
In the third step we take the 
variational parameters of the non-rotating system and modify them to generate an initial guess for the 
rotating system.  We use:        
$
f^{i}_{N_i} \vert_{\text{initial}}=\exp \left(  i \theta_{i} N_i V \right )(1+\eta_{N_i}^{i}) f_{N_i}^{i}\vert_{\Omega=0}, 
$
where the additional variational parameter, $V$, is an integer, $\eta$ is a random complex number, and $\theta_{i}$ is the 
angular coordinate of the site $i$.  The above 
ansatz introduces a vorticity, $V$, while finite $\eta$ ensures that our minimization routine explores 
a variety of minima.  For small systems we obtain identical ground states for all choices of  
$\eta$ configurations.  We conclude that the $\vert \eta \vert=0$ ground state represents a robust minimum.  
For large systems we take $\eta=0$ where convergence is linear in $N$.  We find 
a variety of vortex lattice configurations and mixtures of Mott and superfluid-vortex states depending on parameters.  In 
the following, however, we focus on slow rotation.       

\begin{figure}
\includegraphics[scale=0.35]{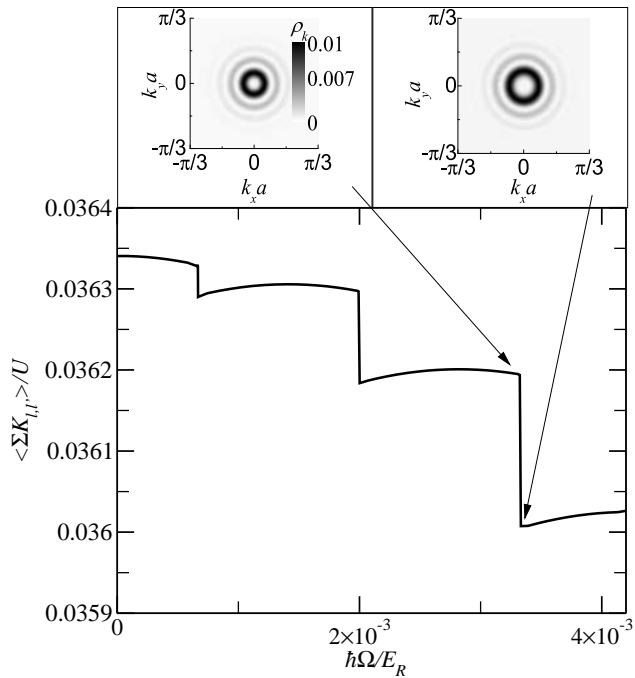}
\caption{Total static kinetic energy of the rotating trapped  
Bose-Hubbard model in the rotating frame plotted versus 
rotation frequency for parameters
($t/U=0.03$ and $\mu/U=0.4$) giving a bulk Mott insulator surrounded by edge 
superfluid.  The steps correspond to increasing vorticity $V=0-3$.  
The top panels plot the momentum distribution in the $k_x k_y$ plane for $\hbar\Omega/E_R=3.31\times10^{-3}$ (left) 
and $3.58\times10^{-3}$ (right).
}
\label{avK}
\end{figure}

We now examine the ground state properties of a rotating system with parameters tuned to give a 
Mott insulator at the trap center with a superfluid strip ($\approx 7$ sites wide) at the edge.  For 
slow rotation (frequencies below the Mott gap) the Mott state rotates with the lattice giving 
zero current in the rotating frame while the superfluid has non-zero current.  
The main panel of Fig.~(\ref{avK}) plots the expectation value of the static
kinetic energy per particle in the rotating frame as a function of the rotation frequency.  The total static 
lattice kinetic energy measures the net change of the ground state phase and therefore 
drops in steps as the superfluid increases vorticity starting from $V=0$.  
The circulation of the edge superfluid jumps when the number of effective flux quanta 
passing through the central Mott insulator increases by an integer to give critical frequencies   
$\Omega_V$ such that:          
$
\hbar \Omega_V /E_R \approx 4V/\left(\pi^2 \vert {\bm r}_{i^e} \vert  ^2 \right).
$
The steps in Fig.~(\ref{avK}) are slightly parabolic because 
the hopping term in $H$ varies as $\Omega^2$:  
$
{\rm Re}\left[ t \exp(iA_{i,j})\right]= t(1+A_{i,j}^2/2+...),
$
which changes the area of the Mott insulator with $\Omega$.

The change in superfluid circulation can be seen in the momentum distribution function and 
may therefore be observable in TOF.  The momentum distribution peaks associated 
with the onset of superfluidity expand stepwise into 
rings of radius $k_r\approx V/(\vert \bm{r}_{i^e} \vert a)$.  
The insets of Fig.~(\ref{avK}) show a grey scale plot 
of $\rho_{{\bm k}}$ in the $k_{x}k_{y}$ plane for two rotation frequencies.  
Here we see that a slight increase of frequency causes a drastic   
change in the shape of the momentum distribution function signaling a jump in 
circulation of the edge superfluid.  Observation of the number of jumps (i.e. $V$), $k_r$, 
and $\Omega_V$ can be used to experimentally overdetermine $\vert {\bm r}_{i^e} \vert$.

\begin{figure}
\setlength{\abovecaptionskip}{-0.3in}
\includegraphics[scale=0.3]{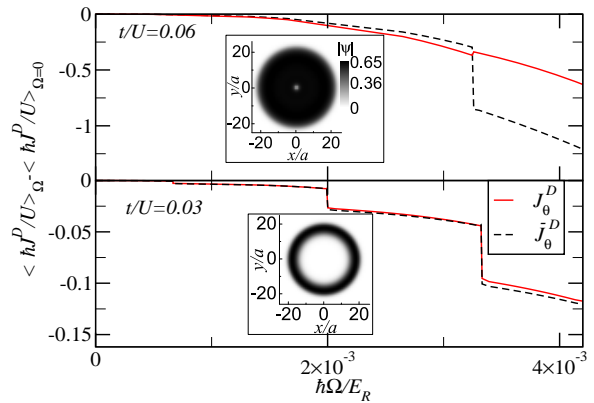}
\caption{ 
Main panels: The diamagnetic current, $J^D_{\theta}$, (solid line) and 
the edge approximation, $\bar{J}^D_{\theta}$, (dashed line) plotted in the rotating frame versus 
rotation frequency.  Insets: The superfluid order parameter plotted in the $xy$ plane.  The top panel 
is entirely in the superfluid phase ($t/U=0.06$ and $\mu/U=0.4$) while the bottom panel is the 
Mott phase with superfluid edges ($t/U=0.03$ and $\mu/U=0.4$).
}
\label{avJ}
\end{figure}

We now ask if the momentum distribution can yield quantitative information 
related to the edge response.  The superfluid rotation in the rotating frame can be 
thought of as a diamagnetic current.  The top panel in Fig.~(\ref{avJ}) plots 
the excess diamagnetic current,  $\langle J^D \rangle_{\Omega}- \langle J^D \rangle_{\Omega=0}$ 
as a function of rotational angular frequency deep in the superfluid regime  
of the trapped BH model with $t/U=0.06$.  The inset shows a grey-scale plot of the 
superfluid order parameter as a function of lattice position 
for $\hbar\Omega/E_R=4.02\times 10^{-3}$.  From the 
plot we see that there is no Mott insulator in the system but there is 
a vortex at the center.  The solid and dashed lines indicate expectation values of 
$J^D$ and $\bar{J}^D$, respectively.  In defining the latter we rewrite the parameter 
$\vert \bm{r}_{i^e} \vert$ in terms of an observable, $\Omega_V$.  The step 
corresponds to the formation of a vortex.  Here we see that the 
approximation made in defining $\bar{J}^D$ does not hold 
for bulk current, i.e. 
$
\langle J_{\theta}^{\rm D} \rangle \not\approx \langle \bar{J}_{\theta}^{\rm D} \rangle. 
$
The superfluid order parameter varies appreciably along the direction transverse to the current 
and, as a result, the diamagnetic current cannot be written 
in the form $\mathcal{J}$.  The bottom panel shows the same but for a different hopping, 
$t/U=0.03$, allowing for a bulk Mott insulator surrounded by edge superfluid (see inset).  
The dashed line reproduces the solid line indicating that  
$
\langle J_{\theta}^{\rm D} \rangle_{e} \approx \langle \bar{J}_{\theta}^{\rm D} \rangle_{e} 
$ 
is in fact a good approximation for an edge superfluid.  Here we find a small spatial 
variance in the superfluid order parameter along the direction 
transverse to the current, i.e. 
$  
\sum_{i}\vert \psi_i \vert^2 (\vert {\bm r_i} \vert-\vert {\bm r_{i^e}} \vert)^2
/\sum_{i}\vert \psi_i \vert^2 \ll 1.
$
We have demonstrated, by a realistic simulation, that one can generate systems 
obeying this small variance condition and that, as a result, the 
edge diamagnetic current can be written in terms of an 
observable, the momentum distribution.  
We propose that in general 
$
\langle J_{\theta}^{\rm D} \rangle 
$ 
can be restored from observation of $\rho_{\bm k}$ and ${\bm k}$ and input parameters to yield a 
powerful tool for studying insulating optical lattice phases with edge states.    
We now discuss potential implications.  

Certain insulators are characterized by a Chern number which can be related to their 
one dimensional edge currents \cite{Hatsugai,Qi}.  
As an example we assume that the Kitaev model \cite{Kitaev,Demler_Spin,Zoller} can 
be realized with two component bosons in a honeycomb optical lattice.  
In Ref.~\cite{Kitaev} 
it was shown that the non-Abelian state can be stabilized in a uniform external magnetic field 
and that edge states exhibit a quantized Righi-Leduc effect.  This prediction asserts 
that the net edge energy-current displays 
a thermal version of the quantum Hall effect where the transverse temperature difference, $T$, between 
the bulk and exterior of the sample establishes a quantized energy current along the edge:
$
J_{{\theta}}^{\rm E}= \nu g (\pi k_B T)^2/6h.
$
Here $g=1$ for bosons and $g=1/2$ for fermions.  In general we expect a clockwise and 
counter-clockwise energy current with an excess number of modes 
$
\nu=\nu_{+\theta}-\nu_{-\theta}.
$  

\begin{figure}
\setlength{\abovecaptionskip}{-0.01in}
\includegraphics[scale=0.3]{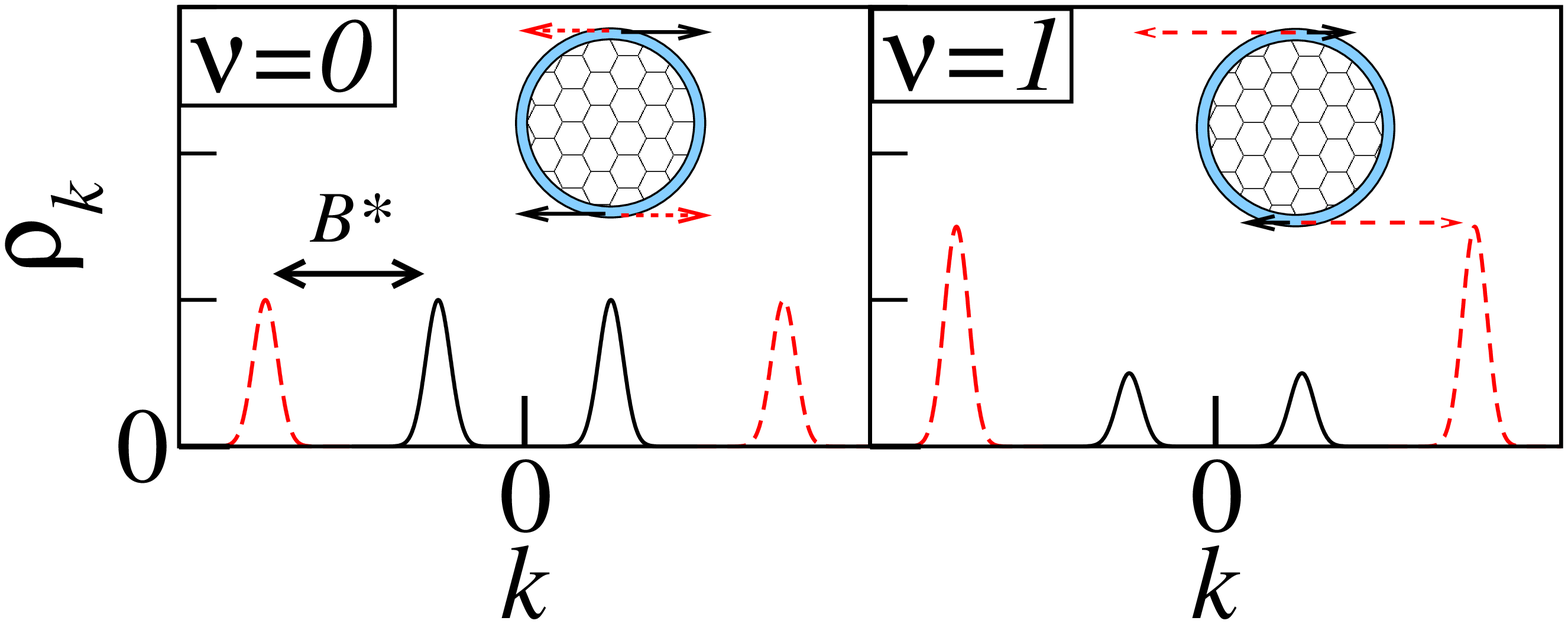}
\caption{Schematic of the expected primary momentum distribution peaks as a 
function of wavevector arising from clockwise and counter-clockwise propagating edge 
superfluids surrounding the Kitaev optical lattice.  The diamagnetic contribution to edge flow 
splits the peaks in an external magnetic field in the Abelian (left) and 
non-Abelian (right) states.
}
\label{schematic}
\end{figure}

We speculate that, in principle, TOF measurements of the momentum distribution function can 
be used to identify chiral edge currents of constituent bosons around two dimensional insulators. 
Flow along the 
edge of the honeycomb lattice ($+\theta$ and $-\theta$) 
corresponds to concentric rings in TOF.  
If $B^{*}$ is chosen to lie at, for example, the $V=1\rightarrow 2$ 
crossing point the $+\theta$ and $-\theta$ modes will 
occupy different momentum channels resulting in 
two concentric rings of differing radii in the momentum distribution 
(Fig.~(\ref{schematic})). 
This suggests that, 
in principle, TOF can be used to study chiral edge current and possibly identify insulators 
with non-zero Chern number.  In practice, however, an observation of edge current in 
TOF pushes current experimental capabilities even for the simplest case of a BH Mott 
insulator.

Sufficiently accurate observations of wave vector and momentum distribution
can be used as a quantitative probe but are difficult to achieve.   
Slow rotation induces only small $k\sim k_r$ modulation of the momentum distribution peaks.  
Small features in the momentum distribution peaks may not be resolvable 
experimentally because TOF measurements are ultimately 
limited in k-space resolution \cite{Spielman}.  
Furthermore, $n_s$ can be adversely affected by interactions during TOF.  Most importantly, 
the number of particles in edge states needs to be sufficiently large to overcome background 
noise in detection.
 
We thank C.W. Zhang for helpful discussion and ARO-DTO, ARO-LPS, and LPS-NSA for support.

\end{document}